\documentclass[10pt,aps,prd,reprint,nofootinbib,floats,floatfix,amsfonts,amssymb,amsmath,preprintnumbers,notitlepage,longbibliography]{revtex4-1}
\usepackage[utf8]{inputenc}
\usepackage[british]{babel}
\usepackage[Symbolsmallscale]{upgreek}
\usepackage{isomath}
\usepackage[protrusion=true,tracking=true,kerning=true,spacing=true,final,babel=true]{microtype}
\usepackage{physics}
\usepackage{xcolor}
\usepackage{graphicx}
\usepackage[final]{hyperref}

\begin{document}
\title{Shot noise in the astrophysical gravitational-wave background}

\author{Alexander~C.~Jenkins}
\affiliation{Theoretical Particle Physics and Cosmology Group, Physics Department, King's College London, University of London, Strand, London WC2R 2LS, United Kingdom}

\author{Mairi~Sakellariadou}
\affiliation{Theoretical Particle Physics and Cosmology Group, Physics Department, King's College London, University of London, Strand, London WC2R 2LS, United Kingdom}

\date{\today}
\preprint{KCL-PH-TH/2019-15}

\begin{abstract}
    We calculate the shot noise induced in the anisotropies of the astrophysical gravitational-wave background by finite sampling of both the galaxy distribution and the compact binary coalescence event rate.
    This leads to a white noise term in the angular power spectrum $C_\ell$, for which we derive a simple analytical expression.
    Failing to account for this term (as previous analyses have done) biases any measurement of the $C_\ell$'s.
    We find that the shot noise dominates over the true astrophysical power spectrum in any reasonable observing scenario, and that only with very long observing times and removal of a large number of foreground sources can the true power spectrum be recovered.
\end{abstract}

\maketitle

\section{Introduction}

The Advanced LIGO~\cite{TheLIGOScientific:2014jea} and Advanced Virgo~\cite{TheVirgo:2014hva} interferometers have instigated an exciting new era of astronomy by directly detecting gravitational waves (GWs).
Eleven detections have been announced thus far~\cite{LIGOScientific:2018mvr}, each associated with a compact binary coalescence (CBC).
However, aside from individual loud events like these, one expects a population of many CBCs that are too faint or numerous to be directly resolvable.
The superposition of GWs from these events leads to the astrophysical gravitational-wave background (AGWB), a persistent GW signal that can be distinguished from instrumental noise by cross-correlating data from multiple detectors~\cite{Romano:2016dpx,Smith:2017vfk}.
The AGWB is a vitally important target for GW observations, and is potentially detectable by LIGO and Virgo at design sensitivity~\cite{Abbott:2017xzg}.

In the literature, the AGWB is often treated as perfectly isotropic, for simplicity.
However, several recent studies~\cite{Jenkins:2018lvb,Jenkins:2018uac,Jenkins:2018kxc,Cusin:2018rsq} have investigated how the anisotropic distribution of GW sources and the inhomogeneous geometry of the spacetime through which the GWs propagate can induce anisotropies in the AGWB.
By carefully studying these anisotropies, one can hope to use the AGWB as a probe of the large-scale structure (LSS) of the Universe, in particular, the distribution of galaxies on large angular scales.

The AGWB anisotropies calculated in Refs.~\cite{Jenkins:2018lvb,Jenkins:2018uac,Jenkins:2018kxc,Cusin:2018rsq} are expressed in terms of the angular power spectrum components $C_\ell$.
These are statistical quantities, describing the expected angular correlation of the AGWB after averaging the signal in at least three distinct ways:
($i$) averaging over random realisations of the cosmic matter distribution (``ensemble of Universes''),
($ii$) averaging over the discrete positions of galaxies within the matter distribution to give a continuous galaxy number density field, and
($iii$) averaging over the merger times of CBCs within each galaxy to give a mean merger rate.
However, the AGWB we observe corresponds to a single Universe, with some discrete number of galaxies, each with some discrete number of CBCs.
In practice, therefore, we are unable to perform the averaging process described above, leading to random fluctuations in the observed $C_\ell$'s, which must be accounted for when comparing to theoretical predictions.
What is more, our most accurate theoretical predictions are themselves based on a simulated galaxy catalogue with a single realisation of LSS and a finite number of galaxies~\cite{Jenkins:2018uac,Jenkins:2018kxc}, so it is doubly important to understand these effects.

References~\cite{Jenkins:2018lvb,Jenkins:2018uac,Jenkins:2018kxc} have already accounted for point ($i$) above: the uncertainty due to our observation of a single realisation of LSS, i.e., cosmic variance.
If the AGWB is Gaussian, then we have the standard result, $\mathrm{Var}\qty[C_\ell]=2C_\ell^2/\qty(2\ell+1)$.
Besides cosmic variance, we must account for the fact that the AGWB is emitted from a finite number of galaxies (sampled from the underlying density field), each hosting a finite number of CBCs (sampled from the mean merger rate).
These two sampling processes, corresponding to points ($ii$) and ($iii$) above, follow Poisson statistics and introduce shot noise to the observed angular power spectrum.
This is a very important effect, which has been studied for decades in contexts such as galaxy redshift surveys~\cite{Feldman:1993ky,Hamilton:2005kz} and the cosmic infrared background~\cite{Kashlinsky:2018mnu}.
In the context of the AGWB, Ref.~\cite{Meacher:2014aca} used numerical simulations to study the effects of shot noise on the monopole (i.e.~isotropic component), but until now the effects of shot noise on the AGWB anisotropies have been ignored.

In this work, we derive expressions for the AGWB angular power spectrum in the presence of shot noise, and calculate the size of the shot-noise effects for realistic models of the AGWB in the LIGO/Virgo frequency band.
We begin with a short description of our model of the AGWB, before defining the angular power spectrum components $C_\ell$, and showing how they are affected by the presence of shot noise.
We then derive an expression for this effect in terms of the CBC population and galaxy distribution, and show how the removal of nearby sources can mitigate it somewhat.
We conclude by discussing the implications of our results for future GW observing runs.

\section{Modelling the AGWB}

Here we briefly recap our model of the AGWB and its anisotropies, first presented in Ref.~\cite{Jenkins:2018uac}.
We describe the AGWB in terms of its density parameter (in units where $c=1$),
    \begin{equation}
        \Omega\qty(\nu_\mathrm{o},\vu*r)\equiv\frac{1}{\rho_\mathrm{c}}\frac{\dd[3]{\rho_\mathrm{gw}}}{\dd{\qty(\ln\nu_\mathrm{o})}\dd[2]{\vu*r}}=\frac{8\uppi G\nu_\mathrm{o}}{3H_0^2}\frac{\dd[3]{\rho_\mathrm{gw}}}{\dd{\nu_\mathrm{o}}\dd[2]{\vu*r}},
    \end{equation}
    which is the energy density of GWs arriving at the observer from direction $\vu*r$, with observer-frame frequency in a logarithmic bin centred on $\nu_\mathrm{o}$, normalised with respect to the critical density of the Universe, $\rho_\mathrm{c}\equiv3H_0^2/\qty(8\uppi G)$.
We treat this as a random field on the sphere, whose angular statistics we wish to study.

The model in Ref.~\cite{Jenkins:2018uac} writes the GW density field as an integral over the population of CBC sources,
    \begin{equation}
    \label{eq:Omega}
        \Omega\qty(\nu_\mathrm{o},\vu*r)=\frac{\uppi}{3}\qty(t_H\nu_\mathrm{o})^3\int\dd{z}\frac{1+z}{E(z)}\int\dd{\vb*\zeta}nR\mathcal{S},
    \end{equation}
    where $t_H\equiv1/H_0$ is the Hubble time, $z$ is the redshift, $E\qty(z)\equiv H\qty(z)/H_0$ is the dimensionless Hubble rate, $\vb*\zeta$ is the set of parameters describing the galaxies and the CBCs, $n\qty(\vb*r,\vb*\zeta)$ is the number density of galaxies with parameter values $\vb*\zeta$ at position $\vb*r$, $R\qty(r,\vb*\zeta)$ is the rate of CBCs per galaxy as a function of redshift, and $\mathcal{S}\qty(\nu_\mathrm{s},\vb*\zeta)$ encodes the GW emission of each CBC as a function of source-frame frequency $\nu_\mathrm{s}$, as given by hybrid waveform models that are calibrated to numerical relativity simulations~\cite{Ajith:2007kx,Ajith:2009bn}.

We calculate the rate function $R$ by convolving the star formation rate of each galaxy with a distribution of delay times (i.e.~the interval between the time of star formation and the coalescence time for a given binary) and normalising to match the local CBC rates inferred by the LIGO/Virgo detections~\cite{LIGOScientific:2018mvr}.
(This calculation also accounts for the suppression of high-mass black hole formation in high-metallicity environments.)
The galaxy number density $n$ is given by a mock galaxy catalogue based on the Millennium Simulation~\cite{Blaizot:2003av,DeLucia:2006szx,Springel:2005nw,Lemson:2006ee}.
This allows us to create a simulated map of the AGWB by individually computing the contribution of each galaxy in the catalogue.
The statistics of this map are then analysed with HEALP\textsc{ix}~\cite{Gorski:2004by}.\footnote{\url{http://healpix.sourceforge.net}}

\section{The angular power spectrum, with and without shot noise}

We now show how the inclusion of generic shot-noise effects in the underlying statistics of the AGWB leads to an additional term in the angular power spectrum.
For later convenience, we write $\Omega$ as an integral over comoving distance,
    \begin{equation}
    \label{eq:Omega-integral-omega}
        \Omega\qty(\nu_\mathrm{o},\vu*r)=\frac{1}{r_H^3}\int\dd{r}r^2\omega\qty(\nu_\mathrm{o},\vb*r),
    \end{equation}
    where $r_H\equiv1/H_0$ is the Hubble radius, and we have defined a new dimensionless function $\omega$.
Comparing with Eq.~\eqref{eq:Omega}, we see that
    \begin{equation}
    \label{eq:omega-definition}
        \omega\qty(\nu_\mathrm{o},\vb*r)=\frac{\uppi}{3}\qty(t_H\nu_\mathrm{o})^3\qty(1+z)\qty(\frac{r_H}{r})^2\int\dd{\vb*\zeta}nR\mathcal{S}.
    \end{equation}
(We suppress the frequency dependence from now on.)

We characterise the anisotropies of $\Omega$ in terms of a multipole expansion of its covariance,
    \begin{equation}
    \label{eq:C_ell-definition}
        C_\ell\equiv\int_{S^2}\dd[2]{\vu*r'}P_\ell\qty(\vu*r\vdot\vu*r')\mathrm{Cov}\qty[\Omega\qty(\vu*r),\Omega\qty(\vu*r')],
    \end{equation}
    which we call the angular power spectrum.\footnote{%
    Note that we do not normalise with respect to the monopole as we did in Refs.~\cite{Jenkins:2018lvb,Jenkins:2018uac,Jenkins:2018kxc}.
    This is because the monopole is now itself a random variable, and accounting for the variance due to this random normalisation would complicate things.
    The physical content of the definition Eq.~\eqref{eq:C_ell-definition} is the same as before.}
Here $\ell$ runs over all non-negative integers, and $P_\ell\qty(x)$ are the corresponding Legendre polynomials.
Note that the $C_\ell$'s are independent of $\vu*r$ due to statistical isotropy.
We can also write Eq.~\eqref{eq:C_ell-definition} as
    \begin{equation}
    \label{eq:Omega_ell_m-cov}
        \mathrm{Cov}\qty[\Omega_{\ell m},\Omega^*_{\ell'm'}]=C_\ell\delta_{\ell\ell'}\delta_{mm'},
    \end{equation}
    where we have decomposed the field into spherical harmonics,
    \begin{equation}
    \label{eq:spherical-harmonics}
        \Omega_{\ell m}\equiv\int_{S^2}\dd[2]{\vu*r}Y^*_{\ell m}\qty(\vu*r)\Omega\qty(\vu*r)=\frac{1}{r_H^3}\int\dd[3]{\vb*r}Y^*_{\ell m}\qty(\vu*r)\omega\qty(\vb*r).
    \end{equation}
Equation~\eqref{eq:Omega_ell_m-cov} shows that each $\Omega_{\ell m}$ is an uncorrelated complex random variable with variance $C_\ell$.

\subsection{The $C_\ell$'s without shot noise}

We calculate the angular power spectrum by specifying the second moment of the $\omega$ field.
Neglecting shot noise, this is simply
    \begin{equation}
    \label{eq:cov-lss}
        \mathrm{Cov}\qty[\omega\qty(\vb*r),\omega\qty(\vb*r')]_\mathrm{LSS}=\bar{\omega}\qty(r)\bar{\omega}\qty(r')\xi\qty(r,r',\theta),
    \end{equation}
    where $\xi$ is the two-point correlation function of $\omega$, describing the probability in excess of random of similar values of the field being clustered together.
Due to statistical isotropy, $\xi$ only depends on the angular positions $\vu*r,\vu*r'$ of the two points through their separation $\theta\equiv\cos^{-1}\qty(\vu*r\vdot\vu*r')$.
(It does, however, depend on their radial distances, as these influence the intensity of the GW flux.)
Using Eqs.~\eqref{eq:Omega-integral-omega},~\eqref{eq:C_ell-definition}, and~\eqref{eq:cov-lss}, we find the $C_\ell$'s in the absence of shot noise,
    \begin{align}
    \label{eq:C_ell-lss}
    \begin{split}
        C_\ell^\mathrm{LSS}&=\frac{1}{r_H^6}\int_{S^2}\dd[2]{\vu*r'}P_\ell\qty(\vu*r\vdot\vu*r')\\
        &\qquad\times\int\dd{r}r^2\int\dd{r'}r'^2\bar{\omega}\qty(r)\bar{\omega}\qty(r')\xi\qty(r,r',\theta).
    \end{split}
    \end{align}

\subsection{The $C_\ell$'s with shot noise}

We now introduce a shot-noise term that encompasses both the galaxy sampling and the CBC rate sampling.
Assuming these effects can jointly be treated as a local Poisson process that is independent of LSS, we have
    \begin{equation}
    \label{eq:cov}
        \mathrm{Cov}\qty[\omega\qty(\vb*r),\omega\qty(\vb*r')]=\bar{\omega}\qty(r)\bar{\omega}\qty(r')\xi\qty(r,r',\theta)+r_H^3\mathcal{V}\qty(r)\delta^3\qty(\vb*r-\vb*r'),
    \end{equation}
    where $\mathcal{V}$ is some function describing the variance due to the finite sample, which is independent of direction due to statistical isotropy.
The factor of $r_H^3$ ensures that $\mathcal{V}$ is dimensionless.

The form of this new shot-noise term [in particular, the fact that it is proportional to $\delta^3\qty(\vb*r-\vb*r')$], is motivated by the equivalent expression for galaxy surveys (see Appendix A of Ref.~\cite{Feldman:1993ky}, or Sec.~2.4 of Ref.~\cite{Hamilton:2005kz}).
It is quite simple to convince oneself that the modification due to shot noise should take this form: there should be an extra term added to the covariance, as shot-noise fluctuations increase the variations in the measured values of $\omega$ throughout space, above the intrinsic variance due to the clustering of GW sources.
However, since the shot-noise fluctuations at one point in space are causally disconnected from those at any other point, the fluctuations at any two points are statistically independent, and the extra term in the covariance should vanish except when the two points are coincident, leading to the delta function.
Equation~\eqref{eq:cov} thus arises naturally from the superposition of independent Poisson processes at each point in space.
We flesh this argument out more quantitatively in Sec.~\ref{sec:calc-shot-power}.

Using Eqs.~\eqref{eq:Omega-integral-omega},~\eqref{eq:C_ell-definition},~\eqref{eq:C_ell-lss}, and~\eqref{eq:cov}, as well as the fact that $P_\ell\qty(1)=1$ for all $\ell$, the full angular power spectrum is then
    \begin{equation}
    \label{eq:lss+W}
        C_\ell=C_\ell^\mathrm{LSS}+\mathcal{W},\qquad\mathcal{W}\equiv\frac{1}{r_H^3}\int\dd{r}r^2\mathcal{V}\qty(r).
    \end{equation}
Thus we see that the shot noise generates a spectrally white (i.e., independent of $\ell$) contribution to the $C_\ell$'s.

\section{Calculating the shot-noise power}
\label{sec:calc-shot-power}

\begin{figure*}[t]
    \includegraphics[width=0.495\textwidth]{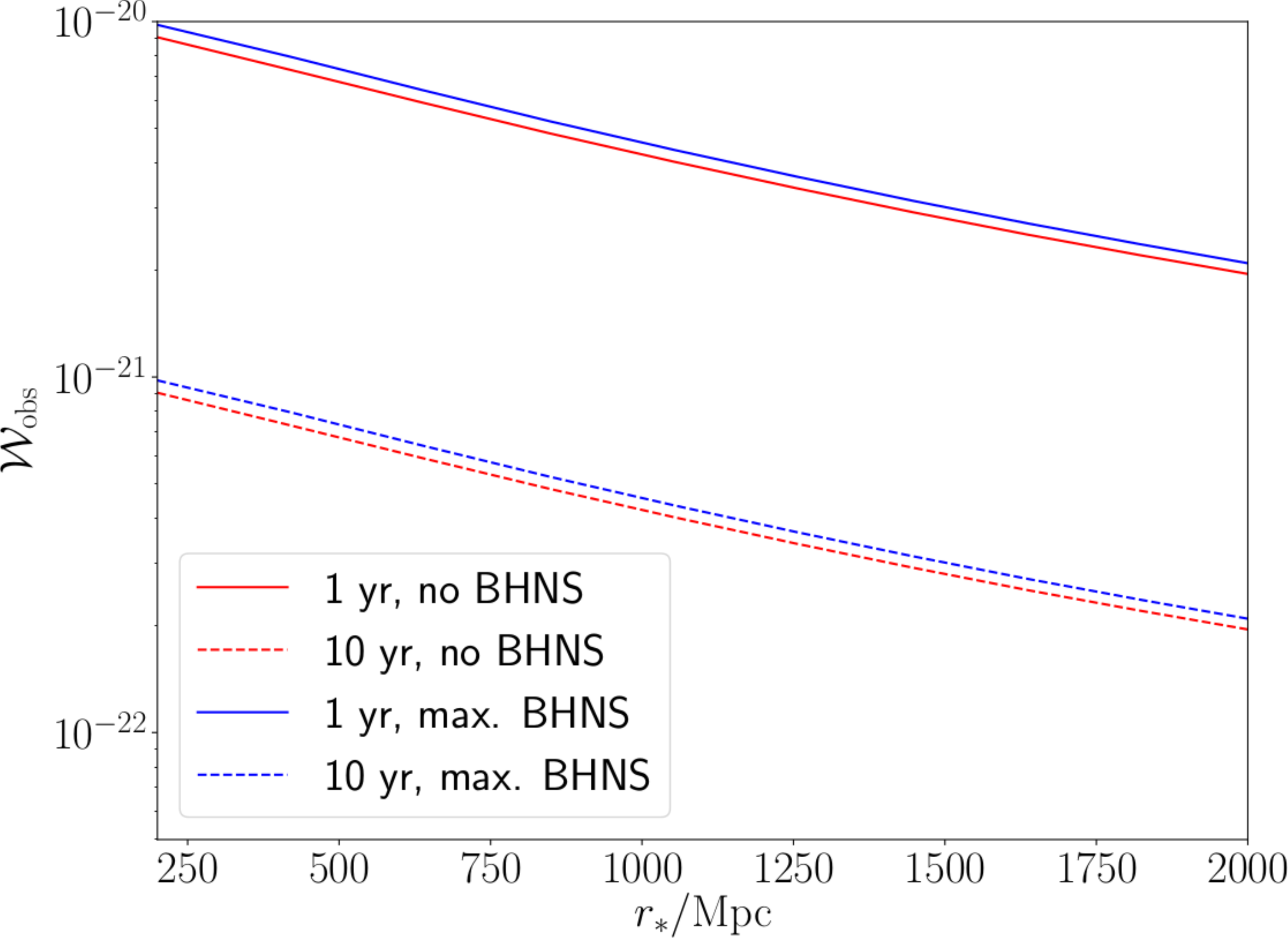}
    \includegraphics[width=0.495\textwidth]{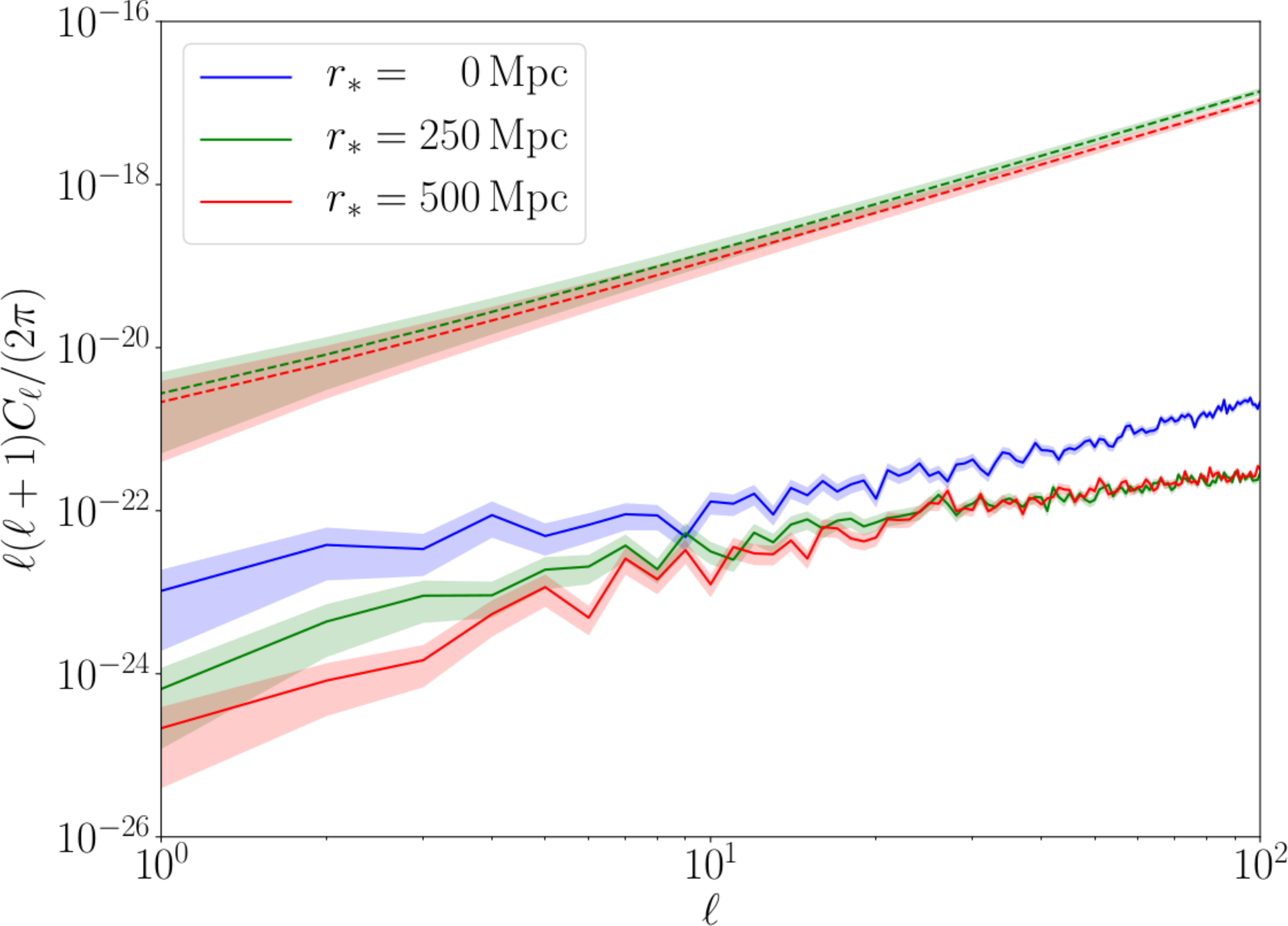}
    \caption{%
        Left panel:
        The observational shot noise $\mathcal{W}_\mathrm{obs}$ as a function of the comoving cutoff distance $r_*$.
        The red curves are calculated using only binary black hole and binary neutron star merger events, at the median rates inferred in Ref.~\cite{LIGOScientific:2018mvr}.
        The blue curves include black hole--neutron star mergers at a rate equal to the upper limit inferred in Ref.~\cite{LIGOScientific:2018mvr}.
        The solid curves represent an observation time $T_\mathrm{o}=1\,\mathrm{yr}$, while the dashed curves represent $T_\mathrm{o}=10\,\mathrm{yr}$, following the scaling $\mathcal{W}_\mathrm{obs}\sim1/T_\mathrm{o}$.
        Right panel:
        The AGWB angular power spectrum, plotted for different values of $r_*$.
        The solid curves are calculated using a mock galaxy catalogue from the Millennium Simulation with the catalogue error $\mathcal{W}_\mathrm{cat}$ removed, and represent the cosmological power spectrum $C_\ell^\mathrm{LSS}$.
        The dashed curves include the observational shot noise after $T_\mathrm{o}=1\,\mathrm{yr}$, giving the full power spectrum $C_\ell=C_\ell^\mathrm{LSS}+\mathcal{W}_\mathrm{obs}$.
        The shaded regions indicate cosmic variance.
        $\mathcal{W}_\mathrm{obs}$ is divergent in the case $r_*=0$, and so is not plotted.
        }%
    \label{fig:W}
    \end{figure*}

In order to evaluate Eq.~\eqref{eq:lss+W}, we must derive an expression for the local Poisson variance function $\mathcal{V}$, accounting for the random sampling of both the galaxy number density and the CBC event rate.

Consider a volume element $\updelta V$ at position $\vb*r$.
We treat the number of galaxies in this region as a Poisson random variable, $N\sim\mathrm{Pois}\qty[\updelta Vn\qty(\vb*r)]$.\footnote{%
We stress that this is only an approximation.
A more sophisticated approach would use the halo model of LSS~\cite{Cooray:2002dia}, accounting for the statistical properties of dark matter halos and of different populations of galaxies within them.
However, this approximation is sufficiently accurate for our purposes (particularly as the galaxy-number contribution to the shot noise is much less than the CBC rate contribution---see below).
}%
Assuming for now that these galaxies and the CBCs in them all have the same parameter values $\vb*\zeta$, the CBC event counts for each galaxy in a given source-frame time interval $T_\mathrm{s}$ are independent and identically distributed Poisson random variables, $\lambda_i\sim\mathrm{Pois}\qty[RT_\mathrm{s}]$.
The total CBC event count from $\updelta V$ is then $\Lambda=\sum_{i=1}^N\lambda_i$,
    which follows a compound Poisson distribution with variance
    \begin{align}
    \begin{split}
        \mathrm{Var}\qty[\Lambda]&=\ev{N}\mathrm{Var}\qty[\lambda]+\ev{\lambda}^2\mathrm{Var}\qty[N]=\updelta Vn\qty(RT_\mathrm{s}+R^2T_\mathrm{s}^2).
    \end{split}
    \end{align}
(This can be shown by computing the moment generating function of $\Lambda$.)
Taking the sampling as independent for different spatial volumes and for different parameter values, we find
    \begin{align}
    \begin{split}
        &\mathrm{Cov}\qty[\Lambda\qty(\vb*r_i,\vb*\zeta),\Lambda\qty(\vb*r_j,\vb*\zeta')]=\updelta V\bar{n}\qty[RT_\mathrm{s}+\qty(RT_\mathrm{s})^2]\delta_{ij}\delta\qty(\vb*\zeta,\vb*\zeta'),
    \end{split}
    \end{align}
    where we have averaged over LSS so that $n\qty(\vb*r)\to\bar{n}\qty(r)$.
Replacing $\Lambda$ with the comoving CBC rate density, $nR\sim\Lambda/\qty(\updelta VT_\mathrm{s})$, and taking $\updelta V\to0$, we are left with
    \begin{equation}
        \mathrm{Cov}\qty[nR,n'R']_\mathrm{shot}=\bar{n}\qty[\frac{R}{T_\mathrm{s}}+R^2]\delta^3\qty(\vb*r-\vb*r')\delta\qty(\vb*\zeta,\vb*\zeta'),
    \end{equation}
    where $n'$ is shorthand for $n\qty(\vb*r',\vb*\zeta')$, etc.
The relationship between the source-frame time $T_\mathrm{s}$ and the observer-frame time $T_\mathrm{o}$ will generally depend on cosmological metric perturbations and peculiar velocities at the location of the source, but to leading order it is simply $T_\mathrm{s}=T_\mathrm{o}/\qty(1+z)$.

Using Eqs.~\eqref{eq:omega-definition} and~\eqref{eq:cov}, the shot-noise power is therefore
    \begin{equation}
    \label{eq:W-main}
        \mathcal{W}=\frac{\uppi^2}{9}\qty(t_H\nu_\mathrm{o})^6\int\dd{r}\qty(\frac{1+z}{r_Hr})^2\int\dd{\vb*\zeta}\bar{n}\qty[\frac{R}{T_\mathrm{s}}+R^2]\mathcal{S}^2.
    \end{equation}
Equation~\eqref{eq:W-main} is our main result, and has two distinct applications: ($i$) interpreting future observations of the AGWB, and ($ii$) improving our theoretical models.
For the former, we can use Eq.~\eqref{eq:W-main} to calculate the expected shot noise in the observed $C_\ell$ spectrum as a function of observing time, given a model for the galaxy number density and CBC merger rate.
The $R/T_\mathrm{s}$ term dominates over the $R^2$ term, as $RT_\mathrm{s}\sim10^{-6}$ for a typical galaxy.
For the latter, we can use Eq.~\eqref{eq:W-main} to calculate the error inherent to our theoretical predictions, due to the finite galaxy sampling in the simulated galaxy catalogue~\cite{Jenkins:2018uac,Jenkins:2018kxc}.
Since these predictions do not involve simulating the time of arrival of discrete GW signals, but average over the CBC rate, they exclude the shot noise due to sampling of this rate.
This is identical to taking the limit $T_\mathrm{o}\to\infty$, meaning that the catalogue predictions contain shot noise due to the $R^2$ term only.
We therefore distinguish between the ``observational shot noise'' and the ``catalogue shot noise,''
    \begin{align}
    \label{eq:W-obs-cat}
    \begin{split}
        \mathcal{W}_\mathrm{obs}&=\frac{\uppi^2\qty(t_H\nu_\mathrm{o})^6}{9T_\mathrm{o}}\int\dd{r}\frac{\qty(1+z)^3}{\qty(r_Hr)^2}\int\dd{\vb*\zeta}\bar{n}R\mathcal{S}^2,\\
        \mathcal{W}_\mathrm{cat}&=\frac{\uppi^2\qty(t_H\nu_\mathrm{o})^6}{9}\int\dd{r}\qty(\frac{1+z}{r_Hr})^2\int\dd{\vb*\zeta}\bar{n}_\mathrm{cat}R^2\mathcal{S}^2.
    \end{split}
    \end{align}
Here, $\bar{n}_\mathrm{cat}$ represents the galaxy number density in the catalogue, which is significantly less than the true galaxy number density $\bar{n}$ (this is because only galaxies brighter than a certain magnitude are included).
Expressing this as a weighted sum of delta functions, with each term representing a single galaxy, the catalogue shot noise is approximated by
    \begin{equation}
        \mathcal{W}_\mathrm{cat}=\sum_i\frac{\uppi\qty(t_H\nu_\mathrm{o})^6\qty(1+z_i)^2}{36r_H^2r_i^4}\int\dd{\vb*\zeta}R_i^2\mathcal{S}_i^2.
    \end{equation}

\section{Removing the foreground}

Inspecting Eq.~\eqref{eq:W-main}, we see that the integrand diverges as $r\to0$.
This is to be expected, for two reasons.
First, the Poisson statistics become progressively worse at small distances, as we are looking at smaller spherical shells that contain fewer galaxies, so the notion of a smooth number density $\bar{n}$ breaks down as $r\to0$.
Second, the contribution of a single CBC to the total AGWB flux becomes much larger at smaller distances, so a CBC that is arbitrarily close can bias the power spectrum by an arbitrarily large amount.

In order to regulate this divergence, it is necessary to introduce a cutoff distance $r_*$, below which we remove any CBC signals and do not consider them part of the AGWB.
(This is similar to what is done with, e.g.,~the cosmic infrared background~\cite{Kashlinsky:2018mnu}.)
We are free to choose the value of $r_*$, with larger values helping to reduce the shot noise as much as possible.
However, the choice of $r_*$ will also affect $C_\ell^\mathrm{LSS}$, and this must be accounted for when making theoretical predictions.

In principle, one can implement this foreground cut by removing from the GW strain time series $h\qty(t)$ any intervals in which an individual CBC with a comoving distance $r<r_*$ is identified.
The CBC chirp signal encodes the luminosity distance $d_L$, which can easily be converted to the comoving distance $r$ by assuming a fiducial cosmology.
However, $d_L$ cannot be measured with arbitrary precision, particularly as it is degenerate with the inclination of the binary.
What is more, the detectability of CBCs at a given distance is also a function of sky position, due to the anisotropic beam pattern of the detectors.
Removing nearby sources based on a signal-to-noise threshold (as in Ref.~\cite{Meacher:2014aca}) risks biasing the power spectrum by favouring particular kinds of CBC events and particular regions of the sky.
It is therefore crucial that $r_*$ is set to be small enough that \emph{all} CBCs at distances $r<r_*$ are detectable, so the foreground removal can be implemented in an unbiased and isotropic manner, without selection effects.
It may be desirable to set different cutoffs for different sources; e.g.,~a much larger value for binary black holes than for binary neutron stars, as the former are detectable at much larger distances.
We leave a detailed examination of these practical issues for future work~\cite{mdc}.

\section{Discussion and conclusion}

Using the AGWB model~\cite{Jenkins:2018uac,Jenkins:2018kxc} described above, we can calculate the size of the observational and catalogue shot-noise terms from Eq.~\eqref{eq:W-obs-cat} for different values of the foreground cutoff distance $r_*$.
We find that the catalogue shot noise is $\mathcal{W}_\mathrm{cat}\approx3\times10^{-29}$ for $r_*=0$, and is typically several orders of magnitude smaller than this for $r_*>0$.
Since the monopole is $\bar{\Omega}\approx10^{-10}$, this represents shot-noise fluctuations of $\lesssim0.01\%$.
This is negligible compared to the true power spectrum $C_\ell^\mathrm{LSS}$, so our earlier predictions in Refs.~\cite{Jenkins:2018uac,Jenkins:2018kxc} are completely unaffected.\footnote{%
In particular, we emphasise that the catalogue shot noise cannot explain why the angular power spectrum calculated in Refs.~\cite{Jenkins:2018uac,Jenkins:2018kxc} is more than an order of magnitude larger than that in Ref.~\cite{Cusin:2018rsq}: the predicted spectrum is all but identical once the catalogue shot noise is subtracted.
We refer the reader to Refs.~\cite{Jenkins:2018uac,Jenkins:2018kxc,Jenkins:2019cau} for discussion of the likely causes of this discrepancy.
    }%

On the other hand, we find that the observational shot noise $\mathcal{W}_\mathrm{obs}$ is generically several orders of magnitude larger than the true angular power spectrum $C_\ell^\mathrm{LSS}$ for any reasonable values of $r_*$ and $T_\mathrm{o}$.
(In fact, $\sqrt{\mathcal{W}_\mathrm{obs}}\approx\bar{\Omega}$, so the shot-noise fluctuations are typically as large as the monopole itself.)
This is illustrated in the right-hand panel of Fig.~\ref{fig:W} for $r_*=250$ and $r_*=500\,\mathrm{Mpc}$.
Note that $C_\ell^\mathrm{LSS}$ also changes with $r_*$, partly because the total GW emission is reduced, causing an overall reduction in the spectrum, and partly because the nearest galaxies contribute most strongly to the AGWB, so that their removal changes the shape of the spectrum.
Calculations of $C_\ell^\mathrm{LSS}$ using the catalogue are not reliable for values of $r_*$ significantly larger than $\sim500\,\mathrm{Mpc}$, due to the incompleteness of the catalogue at high redshift.
However, as can be seen in the left-hand panel of Fig.~\ref{fig:W}, even increasing $r_*$ from $200\,\mathrm{Mpc}$ to $2\,\mathrm{Gpc}$ can only reduce $\mathcal{W}_\mathrm{obs}$ by less than an order of magnitude, so this seems unlikely to solve the problem.

The numerical values given in Fig.~\ref{fig:W} depend on the details of the AGWB model, and include a multitude of random and systematic uncertainties in, e.g., the populations of astrophysical sources that contribute, their emission rates, and the nature of their clustering.
However, we stress that the main result, Eq.~\eqref{eq:lss+W}, is generic, and is grounded in simple and realistic physical principles.
Any finite population of sources will have random Poissonian fluctuations.
If these fluctuations are statistically independent at different spatial locations (i.e., if the shot-noise fluctuations are causally disconnected), then the angular power spectrum generically gains an extra white-noise component, $\mathcal{W}$.
If these sources are finite in time, then basic Poisson statistics dictates that this noise decays as the inverse of the observation time, $\mathcal{W}\propto1/T_\mathrm{o}$.

These results raise the question of whether it is possible to mitigate the shot noise, and thereby probe $C_\ell^\mathrm{LSS}$ on reasonable observational timescales.
In the context of galaxy redshift surveys, there are a range of statistical methods for suppressing the effects of shot noise, many of which involve applying some optimal weighting to each galaxy, such as the ``FKP weighting''~\cite{Feldman:1993ky,Hamilton:2005kz}.
Could similar methods be developed for the AGWB?
The fundamental issue with this idea is that the AGWB consists primarily of unresolved events, and it is therefore not clear how any kind of weighting could be applied in practice.
Note however that in the era of third-generation GW detectors such as the Einstein Telescope~\cite{Punturo:2010zz}, almost every binary black hole coalescence in the observable Universe will be resolvable~\cite{Regimbau:2016ike}, and it may become possible to study LSS with optimally weighted CBC number counts.
Another possibility is to study the cross correlation of the AGWB with electromagnetic (EM) probes of LSS (e.g.~galaxy surveys, or weak lensing).
Since these EM observables have a lower level of shot noise, one would naively expect the shot noise of the cross-correlated spectrum to be smaller than for the AGWB autocorrelation studied here.
This cross-correlated spectrum could be used to test for the presence of primordial black holes in the AGWB, using a similar approach to Ref.~\cite{Scelfo:2018sny}.

The results presented here are of vital importance for interpreting future observations of the AGWB.
They show that it will be much harder than previously thought to uncover the cosmological information encoded in the AGWB anisotropies.
Nonetheless, it remains possible that with sufficient foreground subtraction, and with long enough observing runs, this information might still be accessible by LIGO and Virgo at design sensitivity, or by third-generation GW detectors such as the Einstein Telescope~\cite{Punturo:2010zz}.
This is an important open question that warrants further investigation.

\begin{acknowledgments}
    We thank Vuk Mandic for helpful comments on the manuscript, and for insightful discussions on related issues.
    This work has been assigned the document number LIGO-P1900050.
    The Millennium Simulation databases used in this work and the web application providing online access to them were constructed as part of the activities of the German Astrophysical Virtual Observatory.
    Some of the results in this work have been derived using the HEALP\textsc{ix} package~\cite{Gorski:2004by}.
    A.C.J. is supported by King's College London through a Graduate Teaching Scholarship.
    M.S. is supported in part by the Science and Technology Facility Council (STFC), United Kingdom, under the research grant No. ST/P000258/1.
\end{acknowledgments}

\bibliography{shot-noise}
\end{document}